\documentclass[pdflatex,sn-mathphys-num]{sn-jnl}
\usepackage{graphicx}
\usepackage{amsmath}
\usepackage{hyperref}
\usepackage[most]{tcolorbox}
\usepackage{xcolor}
\usepackage{tikz}
\usetikzlibrary{arrows.meta, positioning}
\usetikzlibrary{arrows.meta, positioning, fit}
\usetikzlibrary{shapes.geometric, arrows.meta, positioning}
\usepackage{pgfplots}
\pgfplotsset{compat=1.18}
\usepackage{tikz}
\usepackage{amsmath}
\usepackage[table]{xcolor}
\usepackage{booktabs}
\usepackage{geometry}
\usepackage{longtable}
\usepackage[utf8]{inputenc}
\usepackage{listings}
\usepackage{xcolor}
\usepackage{tabularx}
\usepackage{enumitem}
\usepackage{amsmath}
\usepackage{amssymb}
\usepackage{lettrine}
\definecolor{queryblue}{HTML}{1F4E78}
\definecolor{responsegray}{HTML}{4C4C4C}
\definecolor{lightgray}{gray}{0.95}

\tcbset{
  on line,
  boxsep=5pt,
  left=0pt,
  right=0pt,
  top=2pt,
  bottom=2pt,
  colframe=black,
  colback=white,
  enhanced,
}

\newtcolorbox{userquerybox}{
  colback=queryblue,
  coltext=white,
  fontupper=\bfseries,
  boxrule=0pt,
  arc=6pt, 
  boxsep=4pt,
  left=4pt,
  right=4pt,
  top=4pt,
  bottom=4pt,
  width=\textwidth,
}

\newtcolorbox{modelresponsebox}{
  colback=responsegray,
  coltext=white,
  fontupper=\bfseries,
  boxrule=0pt,
  arc=6pt, 
  boxsep=4pt,
  left=4pt,
  right=4pt,
  top=4pt,
  bottom=4pt,
  width=\textwidth,
}

\newtcolorbox{responsecontent}{
  colback=lightgray,
  colframe=black,
  coltext=black,
  boxrule=0.5pt,
  arc=6pt, 
  left=8pt,
  right=8pt,
  top=8pt,
  bottom=8pt,
  width=\textwidth,
}
\title{\textbf{CyberLLM-FINDS 2025: Instruction-Tuned Fine-tuning of Domain-Specific LLMs with Retrieval-Augmented Generation and Graph Integration for MITRE Evaluation}
}
\author[1]{\fnm{Vasanth} \sur{Iyer}}  
\author[2]{\fnm{Leonardo} \sur{Bobadilla}}  
\author[2]{\fnm{S. S.} \sur{Iyengar}}  
\affil[1]{\orgdiv{Department of Computer Science}, \orgname{Grambling State University}, \orgaddress{\state{Louisiana}, \country{USA}}}
\affil[2]{\orgname{Florida International University}, \orgaddress{\state{Florida}, \country{USA}}}

\date{\today}

\begin{document}

\maketitle

\begin{abstract}\,Large Language Models (LLMs) such as Gemma-2B have demonstrated remarkable proficiency in various NLP tasks. However, general-purpose models lack deep domain expertise in cybersecurity. This research presents a methodology for fine-tuning the Gemma-2B model into a domain-specific cybersecurity LLM. We outline the dataset preparation, domain fine-tuning process, synthetic data generation, and implications for real-world cybersecurity applications. The results indicate improved translation of threat events from Chain-of-Thought tuning to Instruction-Level tuning within the cybersecurity domain, including threat detection, forensic investigation, and attack analysis.

Further experimentation reveals that domain-specific fine-tuning introduces challenges in prompt length distribution, diverging from patterns in general-purpose models. Uneven prompt lengths complicate the model’s ability to optimize its context window usage, effectively constraining local inference to 200–400 tokens—despite support for 2048. One-shot prompts resembling chain-of-thought reasoning paired with quantized weights performed best. Due to these context window constraints, we employed a hybrid method: cloud-based LLMs generated synthetic datasets, which were then used to fine-tune locally hosted, resource-efficient models.

To extend the evaluation, we introduce a Retrieval-Augmented Generation (RAG) pipeline and graph-based reasoning framework. This enables structured alignment with MITRE ATT\&CK techniques using STIX-based threat intelligence, improving recall in multi-hop and long-context scenarios. The graph modules encode entity-neighborhood context and tactic chains, helping mitigate limitations of short prompt windows. Results show enhanced model alignment with TTP coverage, validating our graph-augmented LLM’s utility in cybersecurity CTI applications.
\end{abstract}

\section{Introduction}
Instruction tuning \cite{yao2023reactsynergizingreasoningacting} is a critical step in adapting large language models (LLMs) like Gemma-2B to domain-specific tasks. It involves fine-tuning the model \cite{Iyer2025-ip},\cite{Iyer2022-mot}\cite{SPIE-2015},\cite{SPIE-2017},\cite{SPIE-2019},\cite{SPIE-2020} on diverse examples framed as natural language instructions across multiple task types. As illustrated in Figure \ref{fig:synthetic-data}, this process enables the model to learn reasoning patterns from tasks such as commonsense inference, translation, and classification, so it can generalize to new tasks like natural language inference—even those it hasn't seen during training. This transferability is essential in cybersecurity \cite{osti_10581950}, where LLMs must respond accurately to structured threats like MITRE ATT\&CK \cite{mitre} techniques without explicit task retraining.

\section{Related Work}
Several works have explored fine-tuning \cite{Iyer2025-ip} large language models for domain-specific applications. Research in medical, legal, and financial domains has shown that adapting LLMs to specialized datasets \cite{Copy-Paste} improves their accuracy. Prior cybersecurity-focused AI models \cite{osti_10581950}, such as OpenAI’s GPT-3 \cite{openai} for threat analysis and IBM Watson for security, highlight the potential of LLMs in this domain. However, challenges such as dataset availability, hallucinations, and security risks remain.

\section{Methodology}
\begin{table}[h!]
\centering
\caption{Size of Popular Language Models (Non-Embedding Parameters)}
\rowcolors{2}{white}{gray!10}
\begin{tabular}{|l|c|c|}
\rowcolor{blue!20}
\textbf{Model} & \textbf{Year} & \textbf{\# Parameters} \\
\rowcolor{blue!20}
 &  & \textbf{(billions, non-embedding)} \\
\hline
BERT Large   & 2018 & 0.34 \\
T5           & 2019 & 11 \\
GPT-3        & 2020 & 175 \\
PaLM         & 2022 & 540 \\
Gemma-2B     & 2023 & 2 \\
DeepSeek-7B  & 2023 & 7 \\
LLaMA        & 2023 & 65 \\
\hline
\end{tabular}
\label{table:foundation-LLM}
\end{table}
Large models typically are trained with lots of data and have many parameters making it challenging to train from scratch in a research environment. So we will explore the current state of the art foundation models which can then be fine tuned to our domain of interest. The criteria for the comparison are the size of the LLMs and context size. Table \ref{table:foundation-LLM} has parameters which are weights and biases ranging from 0.3 billion to 540 billion parameters. The Table \ref{table:foundation-LLM} also shows that the researchers have recently been able to train better models with less number of parameters making it easier for fine tuning for the final tasks. 
In the initial steps we like to optimize memory for efficient tuning and we next evaluate how the number of parameters has an effect on prompting \cite{yao2023reactsynergizingreasoningacting} accuracy as illustrated in Figure \ref{fig:prompting_accuracy_plot} We categorize prompts \cite{yao2023reactsynergizingreasoningacting} as follows: Zero-shot i.e. without training data, one-shot with one example and few-shots with few examples as described in the prompts below. 
\begin{figure}
\centering
\begin{tikzpicture}
\begin{semilogxaxis}[
    width=14cm,
    height=8cm,
    xlabel={Number of Examples in Context (K)},
    ylabel={Accuracy (\%)},
    xmin=0.5, xmax=30,
    ymin=0, ymax=70,
    xtick={1, 2, 5, 10, 20},
    ytick={0, 10, 20, 30, 40, 50, 60, 70},
    legend pos=south east,
    grid=both,
    title={Effect of Prompting Across Model Sizes},
    axis lines=left,
    enlargelimits
]

\addplot[
    color=blue,
    thick
]
coordinates {
    (0.5, 10) (1, 45) (2, 55) (5, 60) (10, 62) (20, 63)
};
\addlegendentry{\tiny 175B Params (Prompt)}

\addplot[
    color=blue,
    dashed
]
coordinates {
    (0.5, 5) (1, 30) (2, 40) (5, 47) (10, 50) (20, 53)
};
\addlegendentry{\tiny 175B Params (No Prompt)}

\addplot[
    color=orange,
    thick
]
coordinates {
    (0.5, 3) (1, 10) (2, 13) (5, 17) (10, 20) (20, 23)
};
\addlegendentry{\tiny 13B Params (Prompt)}

\addplot[
    color=orange,
    dashed
]
coordinates {
    (0.5, 2) (1, 8) (2, 11) (5, 13) (10, 15) (20, 18)
};
\addlegendentry{\tiny 13B Params (No Prompt)}

\addplot[
    color=green!70!black,
    thick
]
coordinates {
    (0.5, 1) (1, 4) (2, 5) (5, 6) (10, 7) (20, 8)
};
\addlegendentry{\tiny 1.3B Params (Prompt)}

\addplot[
    color=green!70!black,
    dashed
]
coordinates {
    (0.5, 0.5) (1, 3) (2, 3.5) (5, 4.5) (10, 5) (20, 6)
};
\addlegendentry{\tiny 1.3B Params (No Prompt)}

\node at (axis cs:0.5,65) [anchor=west] {\small \textbf{Zero-shot}};
\node at (axis cs:1,65) [anchor=west] {\small \textbf{One-shot}};
\node at (axis cs:3,67) [anchor=west] {\small \textbf{Few-shot}};

\node at (axis cs:2.5,57) [anchor=west, rotate=10] {\small Natural Language Prompt};
\node at (axis cs:2.5,42) [anchor=west, rotate=10] {\small No Prompt};

\end{semilogxaxis}
\end{tikzpicture}
\caption{Comparison of model accuracy across parameter sizes and prompting strategies as a function of the number of in-context examples. Prompted models consistently outperform non-prompted models, especially in few-shot regimes.}
\label{fig:prompting_accuracy_plot}
\end{figure}
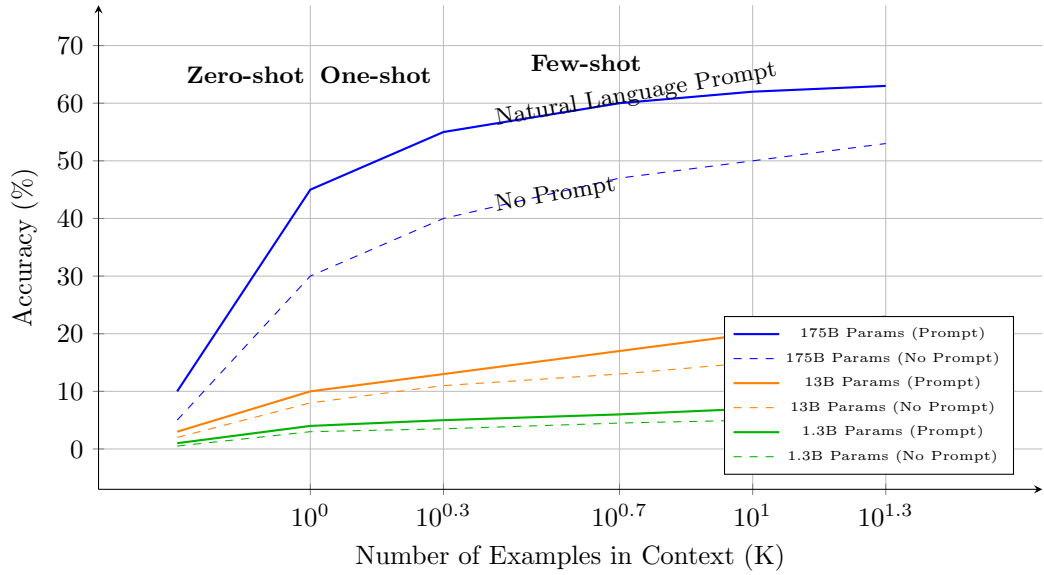
\begin{tcolorbox}[colback=black!5!white, colframe=blue!75!black, title=User Query]
\small Explain Zero-Shot Prompting using MITRE ATT\&CK framework.
\end{tcolorbox}
\begin{tcolorbox}[colback=white, colframe=gray!75!black, title=Model Response]
\textbf{\small Zero-shot prompting} \small means that the model is asked to classify or reason with \textit{\small no prior example} provided.\\
\textbf{Prompt:} \small
\begin{verbatim}
Classify the MITRE ATT&CK technique based on the following activity:
"The attacker executed a PowerShell script to download and run a malicious payload."
Answer:
\end{verbatim}
\textbf{\small Expected Output:} \small
\begin{verbatim}
Tactic: Execution
Technique: T1059 - Command and Scripting Interpreter
Sub-technique: T1059.001 - PowerShell
\end{verbatim}
\end{tcolorbox}
\begin{tcolorbox}[colback=black!5!white, colframe=blue!75!black, title=User Query]
\small Show a one-shot prompting example using MITRE ATT\&CK.
\end{tcolorbox}
\begin{tcolorbox}[colback=white, colframe=gray!75!black, title=Model Response]
\textbf{\small One-shot prompting} \small provides a single example before asking the model to perform a similar task.
\textbf{Prompt:}
\small
\begin{verbatim}
Classify the MITRE ATT&CK technique based on the following activity.
Example:
"The attacker used a phishing email with a malicious attachment to gain initial 
access."
Answer:
Tactic: Initial Access
Technique: T1566 - Phishing
Sub-technique: T1566.001 - Spearphishing Attachment
Now classify this:
"The attacker executed a PowerShell script to download and run a malicious 
payload."
Answer:
\end{verbatim}
\textbf{\small Expected Output:}
\small
\begin{verbatim}
Tactic: Execution
Technique: T1059 - Command and Scripting Interpreter
Sub-technique: T1059.001 - PowerShell
\end{verbatim}
\end{tcolorbox}
\begin{tcolorbox}[colback=black!5!white, colframe=blue!75!black, title=User Query]
Show a few-shot prompting example using MITRE ATT\&CK.
\end{tcolorbox}
\begin{tcolorbox}[colback=white, colframe=gray!75!black, title=Model Response]
\textbf{\small Few-shot prompting} \small provides multiple examples to help the model generalize better.\\
\textbf{Prompt:}
\small
\begin{verbatim}
Classify the MITRE ATT&CK technique based on the following activities.
Example 1:
"The attacker used a phishing email with a malicious attachment to gain initial 
access."
Answer:
Tactic: Initial Access
Technique: T1566 - Phishing
Sub-technique: T1566.001 - Spearphishing Attachment
Answer:
\end{verbatim}
\textbf{\small Expected Output:}
\small
\begin{verbatim}
Tactic: Execution
Technique: T1059 - Command and Scripting Interpreter
Sub-technique: T1059.001 - PowerShell
\end{verbatim}
\end{tcolorbox}
\rowcolors{2}{white}{gray!10}
\begin{longtable}{|p{3.5cm}|c|c|p{6cm}|}
\caption{Summary of locally trainable large language models (LLMs) with fewer than 2 billion parameters. These models are selected based on their ability to fit within standard GPU memory constraints (e.g., 24GB) and support fine-tuning for domain-specific tasks such as cybersecurity reasoning.\label{table:local-LLM}} \\
\hline
\rowcolor{blue!20}
\textbf{Model} & \textbf{Params (B)} & \textbf{Year} & \textbf{Highlights} \\
\hline
\endfirsthead
\hline
\rowcolor{blue!20}
\textbf{Model} & \textbf{Params (B)} & \textbf{Year} & \textbf{Highlights} \\
\hline
\endhead
Gemma-2B & 2 & 2024 & Google’s lightweight open model for on-device and fine-tuning tasks \\
TinyLLaMA-1.1B & 1.1 & 2023 & Minimal resource LLaMA-based model for mobile/IoT research \\
Phi-2 & 2.7 & 2023 & Microsoft’s model designed for reasoning, aligned with on-device use \\
DeepSeek-1.3B & 1.3 & 2024 & DeepSeek’s small model for fast, local inference \\
StableLM-3B & 3 & 2023 & Stability AI’s open model designed for transparency and edge use \\
RedPajama-3B & 3 & 2023 & Open LLaMA-style model trained on reproducible public datasets \\
\hline
\end{longtable}
Figure \ref{fig:prompting_accuracy_plot} illustrates the relationship between prompting accuracy and the number of in-context examples. As the number of examples increases from zero-shot to few-shot, the model's performance improves, making fewer errors and demonstrating better generalization. Additionally, models with larger parameter counts exhibit stronger zero-shot generalization capabilities.

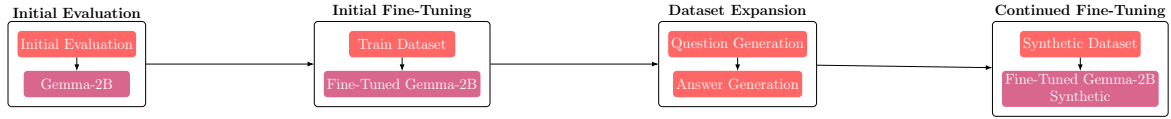
\begin{figure}[ht]
\centering
\resizebox{\textwidth}{!}{%
\begin{tikzpicture}[
  node distance=0.7cm and 4cm,
  every node/.style={align=center},
  task/.style={rectangle, draw=none, rounded corners, minimum width=4cm, minimum height=1cm, fill=red!60, text=white, font=\Large},
  model/.style={rectangle, draw=none, rounded corners, minimum width=4cm, minimum height=1cm, fill=purple!60, text=white, font=\Large},
  groupbox/.style={draw=black, thick, rounded corners, inner sep=10pt},
  arrow/.style={-Latex, thick}
]

\node[task] (eval_label) {Initial Evaluation};
\node[model, below=0.5cm of eval_label] (eval_model) {Gemma-2B};
\node[groupbox, fit=(eval_label)(eval_model), 
      label=above:{\Large\textbf{Initial Evaluation}}] (group1) {};
\draw[arrow] (eval_label) -- (eval_model);

\node[task, right=of eval_label, xshift=4cm] (train_data) {Train Dataset};
\node[model, below=0.5cm of train_data] (ft_model) {Fine-Tuned Gemma-2B};
\node[groupbox, fit=(train_data)(ft_model), 
      label=above:{\Large\textbf{Initial Fine-Tuning}}] (group2) {};
\draw[arrow] (train_data) -- (ft_model);

\node[task, right=of train_data, xshift=4cm] (q_gen) {Question Generation};
\node[task, below=0.5cm of q_gen] (a_gen) {Answer Generation};
\node[groupbox, fit=(q_gen)(a_gen), 
      label=above:{\Large\textbf{Dataset Expansion}}] (group3) {};
\draw[arrow] (q_gen) -- (a_gen);

\node[task, right=of q_gen, xshift=4cm] (syn_data) {Synthetic Dataset};
\node[model, below=0.5cm of syn_data] (syn_model) {Fine-Tuned Gemma-2B \\Synthetic};
\node[groupbox, fit=(syn_data)(syn_model), 
      label=above:{\Large\textbf{Continued Fine-Tuning}}] (group4) {};
\draw[arrow] (syn_data) -- (syn_model);

\draw[arrow] (group1) -- (group2);
\draw[arrow] (group2) -- (group3);
\draw[arrow] (group3) -- (group4);

\end{tikzpicture}
}

\caption{Finetuning LLM}
\label{fig:llm_finetuning}
\end{figure}

\begin{figure}[ht]
\centering
\resizebox{\textwidth}{!}{%
\begin{tikzpicture}[
  node distance=0.7cm and 4cm,
  every node/.style={align=center},
  task/.style={rectangle, draw=none, rounded corners, minimum width=4cm, minimum height=1cm, fill=red!60, text=white, font=\Large},
  model/.style={rectangle, draw=none, rounded corners, minimum width=4cm, minimum height=1cm, fill=purple!60, text=white, font=\Large},
  groupbox/.style={draw=black, thick, rounded corners, inner sep=10pt},
  arrow/.style={-Latex, thick}
]

\node[task] (eval_label) {Initial Evaluation};
\node[model, below=0.5cm of eval_label] (eval_model) {Gemma-2B};
\node[groupbox, fit=(eval_label)(eval_model), 
      label=above:{\Large\textbf{Initial Evaluation}}] (group1) {};
\draw[arrow] (eval_label) -- (eval_model);

\node[task, right=of eval_label, xshift=4cm] (domain_data) {Domain Dataset};
\node[model, below=0.5cm of domain_data] (ft_model) {Fine-Tuned Gemma-2B};
\node[groupbox, fit=(domain_data)(ft_model), 
      label=above:{\Large\textbf{Initial Fine-Tuning}}] (group2) {};
\draw[arrow] (domain_data) -- (ft_model);

\node[task, right=of domain_data, xshift=4cm] (q_gen) {Question Generation};
\node[task, below=0.5cm of q_gen] (a_gen) {Answer Generation};
\node[groupbox, fit=(q_gen)(a_gen), 
      label=above:{\Large\textbf{Dataset Expansion}}] (group3) {};
\draw[arrow] (q_gen) -- (a_gen);

\node[task, right=of q_gen, xshift=4cm] (syn_data) {Synthetic Dataset};
\node[model, below=0.5cm of syn_data] (syn_model) {Fine-Tuned Gemma-2B\\Synthetic};
\node[groupbox, fit=(syn_data)(syn_model), 
      label=above:{\Large\textbf{Continued Fine-Tuning}}] (group4) {};
\draw[arrow] (syn_data) -- (syn_model);

\draw[arrow] (group1) -- (group2);
\draw[arrow] (group2) -- (group3);
\draw[arrow] (group3) -- (group4);

\end{tikzpicture}
}
\caption{Domain Finetuning LLM}
\label{fig:llm_domain_finetuning}
\end{figure}
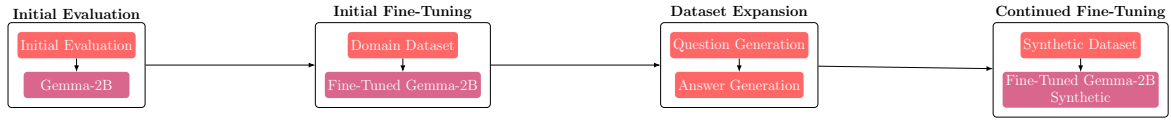

Our training pipeline Figures \ref{fig:llm_finetuning},\ref{fig:llm_domain_finetuning} consists of two main stages: (1) the construction of domain-specific datasets \ref{fig:llm_domain_finetuning} and (2) the fine-tuning of language models. To address resource constraints, we leverage large language models (LLMs) to generate synthetic datasets and utilize smaller, locally runnable LLMs—typically with reduced precision—for fine-tuning. Specifically, we select native LLMs with fewer than 2 billion parameters, as summarized in Table \ref{table:local-LLM}, to ensure compatibility with our available GPU memory. In this work, we focus on domain adaptation using Google’s recently released Gemma-2B model.

\subsection{Dataset Collection}
To develop a cybersecurity-specific expert model, we construct a domain-specific dataset based on the MITRE ATT\&CK framework and use it to fine-tune the language model, as illustrated in Figure~\ref{fig:llm_finetuning}. MITRE ATT\&CK organizes adversarial behavior using well-defined Tactics and Techniques, providing a structured taxonomy that supports effective generalization during fine-tuning. The prompts used for this task are generated using a chain-of-thought prompting approach, enabling the model to reason through sequential steps aligned with the structure of ATT\&CK.
\subsection{Technique T\# in the MITRE ATT\&CK Framework}
This section provides an overview of Technique \#'s as defined in the MITRE ATT\&CK framework, including its associated tactics, use cases, and adversary behaviors. The technique is often leveraged by threat actors to achieve \textit{[specific objective]}.
\subsection{Understanding Vulnerability T\#}
Vulnerability X is a \textit{[type of flaw]} that affects \textit{[systems/applications]}. It allows attackers to \textit{[describe action, e.g., escalate privileges, exfiltrate data, etc.]}. This subsection explains the technical working of the vulnerability, including how it is exploited and its presence in known threat campaigns.

\subsection{Mitigation Strategies for Vulnerability T\#}
To reduce the risk associated with Vulnerability X, organizations can implement several mitigation strategies:
\begin{itemize}
    \item Apply security patches and updates regularly.
    \item Use network segmentation and access controls.
    \item Employ endpoint detection and response (EDR) tools.
    \item Monitor for known indicators of compromise (IoCs).
\end{itemize}
A typical technique and its description is shown in Table \ref{tbl:MITRE-Technique}.
\begin{table}[h!]
\centering
\renewcommand{\arraystretch}{1.3} 
\begin{tabularx}{\textwidth}{@{} >{\bfseries}l @{\hspace{1em}} X @{}}
\toprule
Aspect & Description \\
\midrule
MITRE Technique    & Txxxx – Technique X \\
Vulnerability Type & \textit{e.g., Buffer Overflow} \\
Exploitable By     & \textit{e.g., Remote attackers, malware} \\
Mitigation         & \textit{Patching, EDR, segmentation, etc.} \\
\bottomrule
\end{tabularx}
\caption{Example: Technique and Vulnerability Description.}
\label{tbl:MITRE-Technique}
\end{table}

\section{Synthetic Data Generation}
In cybersecurity, high-quality labeled data is scarce, often sensitive, and typically imbalanced toward benign activity. This presents a significant barrier to effectively fine-tuning large language models (LLMs) for security-specific reasoning tasks. To address this, we introduce a synthetic data generation \cite{Iyer2025-ip} framework as illustrated in Figure~\ref{fig:synthetic-data}. By leveraging the structured nature of the MITRE ATT\&CK framework, we use LLMs to generate instruction-style examples that simulate a wide variety of attack tactics, techniques, and procedures (TTPs). This pipeline enables the creation of logs \cite{zeek}, \cite{ibm} for rare or hard-to-collect threats, supports diverse prompting styles (e.g., zero-shot, one-shot, few-shot), and allows for balanced datasets that improve fine-tuning efficiency and model generalization. Crucially, it also provides a privacy-preserving and legally compliant alternative to real-world security logs.
\subsection{Instruction Tuning Process}
We perform fine-tuning across multiple tasks at the instruction level.
\begin{figure}
    \centering
    \includegraphics[scale=0.9]{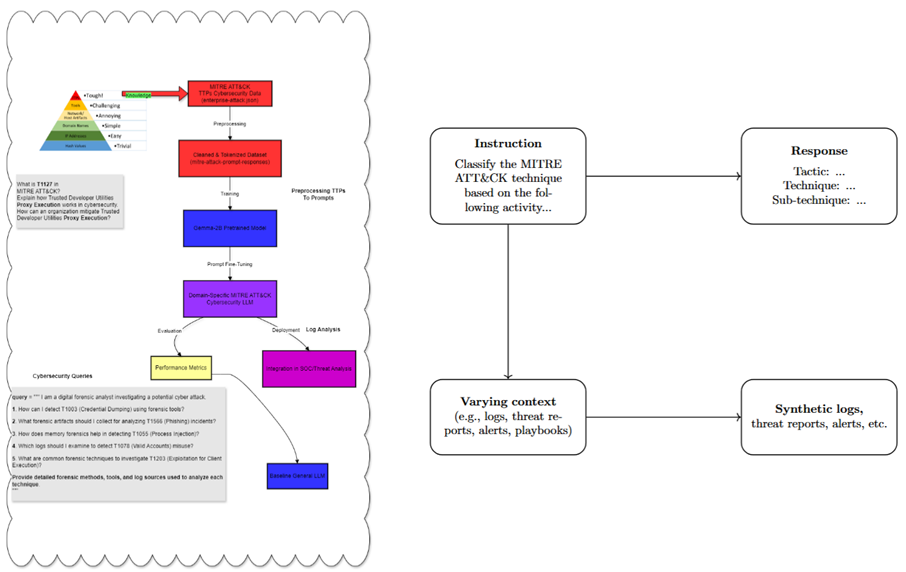}
    \caption{ Workflow for generating synthetic cybersecurity data using large LLMs. The process produces instruction-format samples that encode attack behaviors, supporting Chain-of-Thought prompting and enabling small models to reason over structured threat intelligence such as MITRE’s Pyramid of Pain, which ranks the difficulty of detecting and disrupting various attacker artifacts.}
    \label{fig:synthetic-data}
\end{figure}
Instruction tuning enables large language models (LLMs) to specialize in domain-specific reasoning by learning from natural language examples aligned with real-world tasks. In the context of cybersecurity, we leverage structured knowledge from the MITRE ATT\&CK framework—such as techniques like T1059 and T1547, tactics including Execution, Persistence, and Lateral Movement, and contextual formats like logs, alerts, threat reports, and playbooks—to construct a diverse and targeted training corpus. As illustrated in Figure~\ref{fig:synthetic-data}, our approach combines domain adaptation with synthetic data generation to support instruction-level fine-tuning at scale.

Synthetic datasets offer the advantage of complete control over coverage and balance. They allow us to generate labeled examples for every ATT\&CK technique, including rare or underrepresented behaviors that are seldom encountered in enterprise environments. For instance, we can simulate advanced scenarios such as \textbf{T1003.001 – LSASS Dumping} using PowerShell-based indicators that may not naturally appear in historical logs. This ensures comprehensive coverage across tactics, sub-techniques, and platforms, including Windows, Linux, and macOS.

Additionally, synthetic logs as in Table \ref{tab:synth_table_label} can be tailored to varying levels of complexity. We design some examples with clean and distinct attack signatures to support basic classification tasks, while others contain obfuscated patterns or mixed signals to train models for reasoning under uncertainty. This form of data augmentation enables models to engage in chain-of-thought prompting—reasoning through multi-step sequences to correctly identify attacker behavior and map it to a specific tactic or technique.

An equally important benefit of using synthetic data is its safety and compliance. Since no real personal or organizational identifiers are involved, this method avoids the legal and ethical risks associated with handling real logs, such as violations of privacy regulations like GDPR or HIPAA. Furthermore, it mitigates the model’s dependence on the biases of any single SOC dataset, promoting better generalization.

Finally, our framework supports the generation of diverse prompt formats for instruction tuning. We create examples suitable for zero-shot learning (where no prior examples are given), one-shot prompts (with a single reference example), and few-shot configurations (featuring multiple labeled examples followed by a query). This variety improves the model’s ability to generalize across different log structures and behavioral patterns. Together, these design choices result in a balanced and task-relevant training set that significantly enhances the model’s performance in detecting and reasoning about cyber threats.

\begin{table}[]
    \centering
\begin{tabular}{@{} >{\raggedright\arraybackslash}p{4.5cm} >{\raggedright\arraybackslash}p{5.5cm} >{\raggedright\arraybackslash}p{5.5cm} @{}}
\toprule
\textbf{Synthetic Log} & \textbf{Instruction} & \textbf{Model Output} \\
\midrule
\texttt{"2024-04-15 10:22:11" user: SYSTEM ran: "powershell -enc ..."} & What MITRE ATT\&CK technique does this log indicate? & Tactic: Execution \newline Technique: T1059.001 – PowerShell \\
\addlinespace
\texttt{Zeek conn.log: 192.168.1.100 → 10.0.0.10 TCP 3389} & Explain what this log suggests and map it to MITRE ATT\&CK. & Indicates use of RDP for Lateral Movement. \newline Tactic: Lateral Movement \newline Technique: T1021.001 – Remote Desktop Protocol \\
\bottomrule
\end{tabular}
\caption{ATT\&CK Prompts Enhanced with Synthetic Data Logs.}
    \label{tab:synth_table_label}
\end{table}
\subsection{Model Evaluation}
The model was evaluated on:
\begin{itemize}
    \item Accuracy in answering MITRE ATT\&CK queries
    \item Performance on cybersecurity question-answering tasks
    \item Effectiveness in analyzing threat logs
    \item Comparison with general-purpose LLMs
\end{itemize}
The first of the four evaluation criteria pertains to domain adaptation. Our 2B-parameter model demonstrated effective fine-tuning and successfully answered Chain-of-Thought-style queries aligned with the MITRE ATT\&CK framework. In contrast, the remaining three criteria—focused on instruction tuning and its extension through synthetic data generation—exhibited lower performance. This was primarily due to the baseline accuracy of models under 2 billion parameters, which remained below 20\%, as illustrated in Figure~\ref{fig:prompting_accuracy_plot}. Due to reduced task accuracy and inadequate alignment with instruction-based tasks, we leveraged larger, cloud-hosted LLMs exceeding 175 billion parameters to generate synthetic instruction datasets. These models, achieving over 50\% accuracy in instruction-following tasks, exhibited stronger generalization capabilities. The generated synthetic data was then used to re-train the smaller, local models \cite{Update-qlora2023}, enhancing their ability to respond to instruction-level prompts.
\section{Results and Discussion}
\subsection{RAG vs Graph-based Retrieval Evaluation on MITRE Queries}

To assess how retrieval augmentation impacts fine-tuned CyberLLMs for MITRE ATT\&CK-style queries, we conducted a detailed comparison between pure RAG, Graph+LLM, and GraphRAG+GNN pipelines. These methods were tested using an automated LLM Judge scoring system across 5 dimensions: relevance, completeness, accuracy, specificity, and clarity. The scripts and dataset used are available in Github\cite{mitre-gnn-analysis-2026}.

\begin{table}[ht]
\centering
\caption{LLM Judge Evaluation for 5 MITRE ATT\&CK Queries}
\label{tab:rag-graph-results}
\begin{tabular}{@{}lcccccc@{}}
\toprule
\textbf{Approach} & \textbf{Relevance} & \textbf{Complete.} & \textbf{Accuracy} & \textbf{Specif.} & \textbf{Clarity} & \textbf{Avg. Score} \\
\midrule
Pure RAG         & 8.2 & 7.9 & 7.8 & 7.6 & 8.0 & \textbf{7.87} \\
Graph + LLM      & 7.5 & 7.0 & 7.3 & 6.8 & 7.2 & 7.16 \\
GraphRAG + GNN   & 8.0 & 7.8 & 8.2 & 7.9 & 8.1 & \textbf{8.00} \\
\bottomrule
\end{tabular}
\end{table}

\noindent The results demonstrate that the hybrid GraphRAG+GNN architecture outperforms pure Graph traversal while matching or exceeding RAG in several categories. Specifically:
\begin{itemize}
    \item \textbf{GraphRAG+GNN} achieved the highest overall score (8.00), with improvements in Accuracy and Specificity.
    \item \textbf{Pure RAG} led in Clarity and speed, winning 3 out of 5 head-to-head evaluations.
    \item \textbf{Graph+LLM} trailed slightly in average score due to broader but less focused responses.
\end{itemize}

These findings suggest that combining structured graph context with lightweight GNN-based node scoring enhances LLM interpretability and retrieval quality on MITRE-style queries.

\begin{figure}
    \centering
    \includegraphics[scale=0.8]{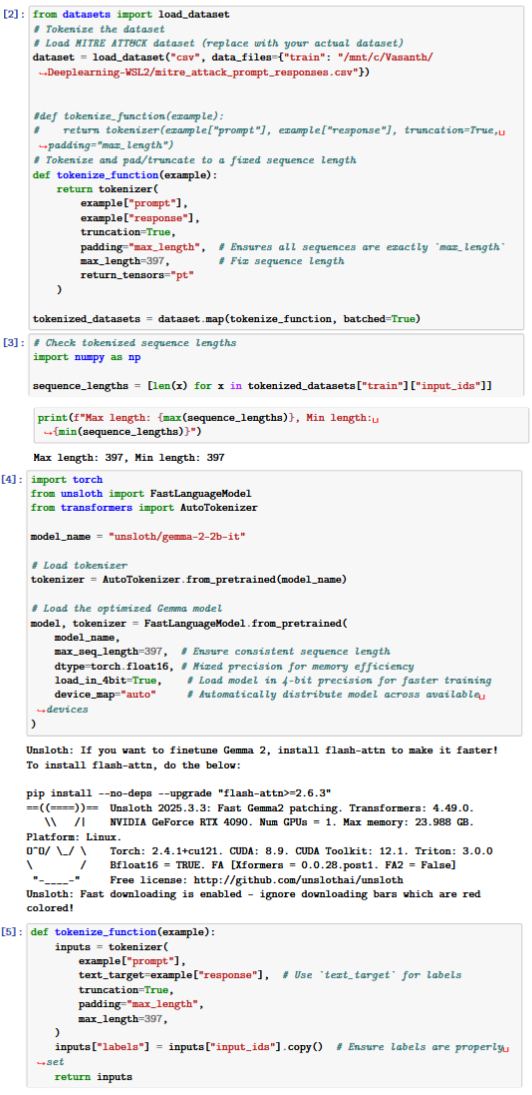}
    \caption{Instruction-tuning workflow using domain-adapted synthetic data. The pipeline integrates structured prompts derived from the MITRE ATT\&CK framework and Chain-of-Thought prompting strategies. The model is fine-tuned with a mixture of zero-shot, one-shot, and few-shot examples to support reasoning and classification across cybersecurity tasks.}
    \label{fig:MITRE-training-code-label}
\end{figure}

Table~\ref{tab:token-training-llms} summarizes viable local LLM training configurations in Figure \ref{fig:MITRE-training-code-label} using current-generation NVIDIA hardware. In our experiments, we utilized a 24GB GPU\textsuperscript{*} for fine-tuning. The domain-specific MITRE ATT\&CK dataset consisted of 2,398 prompt-response pairs (\ref{fig:MITRE-training-code-label}). To accommodate GPU memory constraints, the batch size was set to 4. However, attempts to use the default token lengths of 1,024 to 2,048 tokens resulted in parser errors at this batch size. We found that a token length of 397 (\ref{fig:MITRE-training-code-label})—combined with dynamic padding for variable-length prompts—enabled stable training across all epochs. This constraint, however, significantly limited training to 1–2 shot prompting scenarios.

\begin{figure}
    \centering
    \includegraphics[scale=0.8]{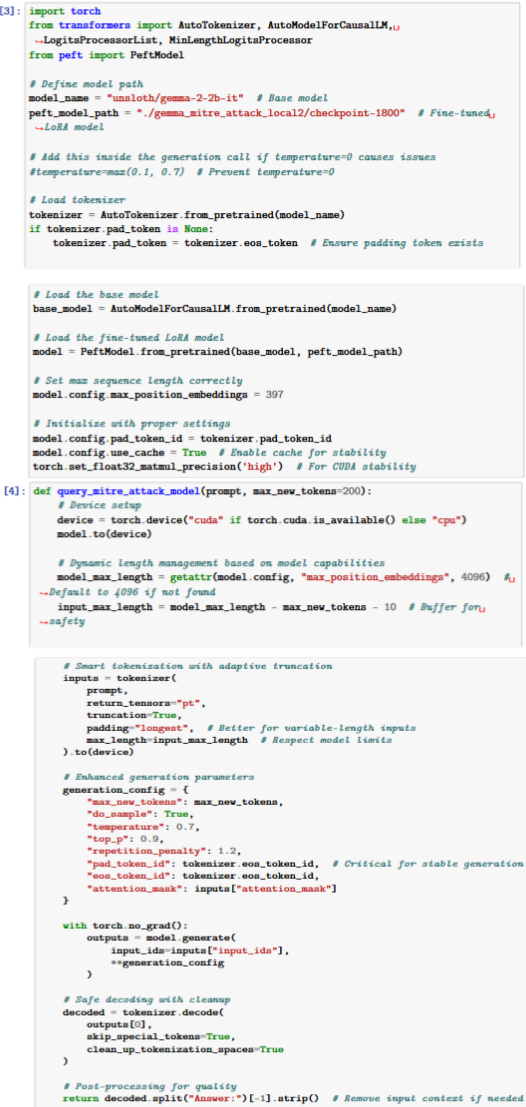}
    \caption{Model inference results for domain-specific instruction prompts. Each output is constrained to a maximum of 200 tokens, reflecting the token limit applied during inference for consistency across prompt evaluations. The responses demonstrate how the fine-tuned model interprets MITRE ATT\&CK-aligned queries, illustrating task comprehension and reasoning within the token constraint.}
    \label{fig:loading-weights-label}
\end{figure}

\begin{figure}
    \centering
    \includegraphics[scale=0.8]{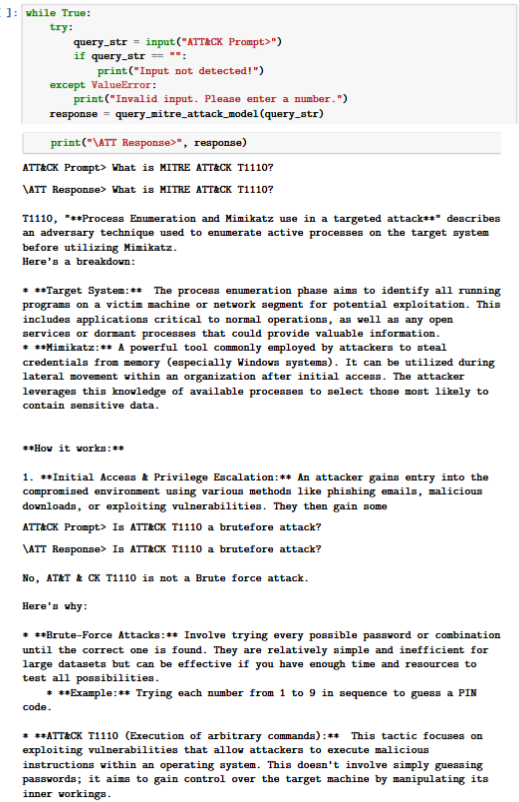}
    \caption{Example of instruction-level fine-tuning with 1–2 shot prompting using synthetic data. The prompt contains a real-world cybersecurity scenario aligned to MITRE ATT\&CK, followed by a response from the fine-tuned model. This illustrates the model's ability to generalize and explain attack techniques based on few-shot learning with padded token limits.}
    \label{fig:inferencecode-label}
\end{figure}

The results of 1–2 shot prompting are presented in Figure~\ref{fig:inferencecode-label}, with the maximum token output length configured to $200$ tokens, as shown in Figure~\ref{fig:loading-weights-label}. The generated responses demonstrate meaningful domain adaptation and interpretable accuracy in the cybersecurity context. However, to comprehensively assess model performance, further evaluation is required. In future work, we plan to compare the fine-tuned model against other standard LLM baselines using an automated LLM-based judge framework to ensure consistent and objective scoring across tasks.
\begin{table}[h!]
\centering
\renewcommand{\arraystretch}{1.4}
\begin{tabular}{@{} p{4cm} p{5cm} p{5cm} @{}}
\toprule
\textbf{GPU Memory Range} & \textbf{Typical Training Config} & \textbf{Notes / Recommendations} \\
\midrule
\textbf{8–16GB GPUs} \newline (e.g., RTX 3060, 4060, 3090, T4) & 
\begin{itemize}
    \item Token limit: \textbf{2048–3072}
    \item Batch size: \textbf{2–4}
    \item Use \texttt{fp16} or \texttt{bf16}
\end{itemize}
& 
\begin{itemize}
    \item Enable \textbf{gradient checkpointing}
    \item FlashAttention can reduce memory cost
    \item Ideal for instruction tuning with 1–2 shot prompts
\end{itemize}
\\
\midrule
\textbf{24–32GB GPUs} \newline (e.g., RTX 4090, A5000, V100) & 
\begin{itemize}
    \item Token limit: \textbf{4096–8192}
    \item Batch size: \textbf{4–8+}
    \item Supports long CoT prompts
\end{itemize}
& 
\begin{itemize}
    \item Can fine-tune with \textbf{4–6 shot} examples
    \item Ideal for multi-turn logs or threat reasoning
    \item Combine synthetic logs with multi-step output
\end{itemize}
\\
\midrule
\textbf{24GB GPUs}$^{*}$ \newline (e.g., RTX 4090) & 
\begin{itemize}
    \item Token limit: \textbf{397}
    \item Batch size: \textbf{4}
    \item Used 4-bit quantized \cite{Update-qlora2023} weights and 16-bit arithmetic
\end{itemize}
& 
\begin{itemize}
    \item Can fine-tune with \textbf{1–2 shot} examples
    \item Ideal for domain fine-tuning locally
\end{itemize}
\\
\bottomrule
\end{tabular}
\caption{$^{*}$Configuration shown was used during both training and evaluation.}
\label{tab:token-training-llms}
\end{table}
\section{Acknowledgments}
Support for this research was provided by the Army Research Office under Grant Number W911NF-21-1-0264. The authors would like to thank Dr. Igor Ternovskiy, for valuable discussions on the use of synthetic data to better approximate domain-specific distributions for fine-tuning large language models. Additional support was provided by the National Science Foundation under Grant Number HBCU-EiR-2101181 and DOE Building Training and Assessment Centers Grants Program for work on developing AI deep learning techniques using explainable AI. Portions of this work also contributed to the development of introductory AI courses, supported by a grant from the Google TensorFlow team.

We gratefully acknowledge the contributions and mentorship of our former Co-Principal Investigator, Dr. Y.B. Reddy, whose guidance and dedication were instrumental in the early stages of this research. Dr. Reddy passed away recently, and we respectfully dedicate this work to his memory.

\end{document}